\documentclass[11pt]{article}
\usepackage{graphicx}

\newcommand{ \be}{\begin{equation}}
\newcommand{ \ee}{\end{equation}}
\newcommand{ \bea}{\begin{eqnarray}}
\newcommand{ \eea}{\end{eqnarray}}
\newcommand{ \mysmall}[1]{\scriptscriptstyle #1} % a smaller #
\newcommand{ \smallmax}{{\rm\scriptstyle max}} 
 
\newcommand{ \mw}{M_{\mysmall{W}}}

\newcommand{ \gl}{g_{\mysmall{L}}}
\newcommand{ \gr}{g_{\mysmall{R}}}

\newcommand{ \nuov}{\bar{\nu}}
\newcommand{ \Aov}{\bar{A}}
\newcommand{ \Bov}{\bar{B}}
\newcommand{ \Cov}{\bar{C}}

\newcommand{ \fig}[1]{fig.~\ref{figure:#1}}
\newcommand{ \eq}[1]{eq.~(\ref{eq:#1})}

\newcommand{ \gev}  {\mbox{ GeV}}
\newcommand{ \mev}  {\mbox{ MeV}}
\newcommand{ \kev}  {\mbox{ KeV}}

\newcommand{ \psl}  {p \kern-.45em{/}}
\newcommand{ \qsl}  {q \kern-.45em{/}}

\newcommand{\BeSeven}{$\!\!\!\!\phantom{A}^7$Be~}

\begin{document}

%--------------------------  TITLE PAGE  --------------------------------%
\begin{titlepage}
\begin{flushright}
%        \small
        BUTP 2002/01\\
        % hep-ph/00xxxxx \\
        % \today
	January 2002
\end{flushright}
\renewcommand{\thefootnote}{\fnsymbol{footnote}}

\begin{center}
\vspace{1cm}
{\LARGE \bf Elastic Scattering of Neutrinos off 
Polarized Electrons\footnote{Work partially supported by 
Schweizerischer Nationalfonds.}}

\vspace{1cm}
{\large\bf      P.~Minkowski$^a$  and 
                M.~Passera$^{a,b}$}

\setcounter{footnote}{0}
\vspace{.8cm}
{\it    $^a$ Institut f\"{u}r Theoretische Physik,
        Universit\"{a}t Bern \\ 
        Sidlerstrasse 5, CH-3012, Bern, Switzerland \\
	Email: {\rm mink@itp.unibe.ch} \\
\vspace{3mm}
        $^b$ Dipartimento di Fisica ``G.~Galilei'', Universit\`{a} 
        di Padova and \\ INFN, Sezione di Padova, Via Marzolo 8,
        I-35131, Padova, Italy \\
	Email: {\rm passera@pd.infn.it}}

\vspace{2cm}
{\large\bf Abstract} 
\end{center}

\vspace{3mm} 
\noindent 
We calculate the cross sections for elastic $\nu_l + e^- \rightarrow
\nu_l + e^-$ and $\nuov_l + e^- \rightarrow \nuov_l + e^-$ scattering
($l=e$, $\mu$ or $\tau$) in the Born approximation and with exactly
fixed polarization states of target and final-state electrons,
discussing their sensitivity to the incident (anti)neutrino flavor. We
suggest investigation of the flavor composition of a (anti)neutrino
beam by a flux-independent analysis of the scattering of its
constituents off polarized electrons.

\end{titlepage}

%-----------------------------   TEXT  -------------------------------------%
%2345678901234567890123456789012345678901234567890123456789012345678901234567
%        1         2         3         4         5         6         7

\noindent The main objective of this letter is to present analytic
formulae for the differential cross sections of the elastic
scatterings $\nu_l + e^- \rightarrow \nu_l +e^-$ and $\nuov_l +e^-
\rightarrow \nuov_l+e^-$ ($l=e$, $\mu$ or $\tau$), and to propose
investigation of the flavor of (anti)neutrinos by a flux-independent
analysis of their elastic scattering off polarized electrons. We
investigate these scattering processes to lowest order in the
electroweak interaction with all polarization states specified. We
neglect terms of order $r/\mw^2$, where $r$ indicates any of the
Mandelstam variables intervening in the scattering process and $\mw$
is the $W$ boson mass. Neutrinos and antineutrinos are considered
massless.

Consider the frame of reference in which the electron is initially at
rest and let the initial electron polarization vector lie in an
arbitrary direction {\boldmath{$\hat{n}$}}.  If we choose helicity to
describe the polarization of the final-state electron, the Standard
Model (SM) prediction for the elastic neutrino--electron differential
cross section is
\be
   \left[\frac{d\sigma}{dE}\right]^{\hat{n},\pm}_\nu 
   \! = \; \frac{2mG_{\mu}^2}{\pi}
   \left[\,\gl^2 A_\pm(E) 
   +\gr^2 B_\pm(E)
   \left(1-z\right)^2 -\gl \gr C_\pm(E) 
        \frac{m z}{\nu}\right],
\label{eq:fullpolarNu}
\ee
where
\bea
	A_\pm(E) &=& \frac{1}{2}\left(1+c_\alpha \right) 
		     \left( 1\mp \cos\theta \right)	\nonumber \\
		%  =   \frac{1}{2}\left(1+c_\alpha \right)
		%\left[ 1 \mp \left(1+\frac{m}{\nu}\right)
		%\frac{T}{l} \right] 			\nonumber \\
	B_\pm(E) &=& \left[ \frac{1}{2} \left(1-c_\alpha \right) 
		+ \frac{m z c_\alpha}{2(\nu-T)} \right]
		\left[1\pm \left(1+\frac{m}{T-\nu}\right)
		\frac{T}{l}\right]			\nonumber \\
	C_\pm(E) &=& \frac{1}{2}\left(1+c_\alpha \right) 
		\left( 1 \pm \frac{T-2\nu}{l} \right).	\nonumber
\eea
The upper (lower) sign indicates positive (negative) helicity of the
recoil electron and $c_\alpha$ is the cosine of the angle between
{\boldmath{$\hat{n}$}} and the incoming neutrino
momentum. $G_{\mu}=1.16637(1) \times 10^{-5}\gev^{-2}$ is the Fermi
coupling constant, $m$ is the electron mass, $\gl = \sin^2
\!\theta_{\mysmall{W}} \pm 1/2$ (upper sign for $\nu_e$ and $\nuov_e$,
lower sign for $\nu_{\mu,\tau}$ and $\nuov_{\mu,\tau}$), $\gr =
\sin^2\!\theta_{\mysmall{W}}$, and $\sin^2\!\theta_{\mysmall{W}}
\approx 0.23$ is the squared sine of the weak mixing angle. In this
elastic process, $E$, the electron recoil energy, ranges from $m$ to
$E_{\smallmax} =$ $[m^2 +(2\nu +m)^2]/[2(2\nu +m)]$, where $\nu$ is
the incident neutrino energy in the frame of reference in which the
electron is initially at rest. Also, $l=\sqrt{E^2-m^2}$ and $T=E-m$
are the final electron three-momentum and kinetic energy, $z=T/\nu$,
and $\cos\theta = (1+m/\nu)(T/l)$ is the cosine of the angle between
the momenta of the recoil electron and incident neutrino. For
simplicity of notation, only the $E$ dependence of the functions
$A_\pm$, $B_\pm$ and $C_\pm$ has been explicitly indicated. 
Note that the differential cross section in \eq{fullpolarNu} includes
an integration over the nontrivial dependence on the azimuthal angle
of the final-state electron momentum, with the $z$-axis taken along the
direction of the incoming neutrino momentum.

The corresponding formula for the elastic antineutrino--electron
differential cross section is
\be
   \left[\frac{d\sigma}{dE}\right]^{\hat{n},\pm}_{\nuov} 
   \! = \; \frac{2mG_{\mu}^2}{\pi}
   \left[\, \gl^2 \Aov_\pm(E) \left(1-z\right)^2  
   +\gr^2 \Bov_\pm(E) -\gl \gr \Cov_\pm(E) 
        \frac{m z}{\nu}\right],
\label{eq:fullpolarANu}
\ee
where
\bea
	\Aov_\pm(E) &=& \left[ \frac{1}{2} \left(1+c_\alpha \right) 
		- \frac{m z c_\alpha}{2(\nu-T)} \right]
		\left[1\mp \left(1+\frac{m}{T-\nu}\right)
		\frac{T}{l}\right]			\nonumber \\
	\Bov_\pm(E) &=& \frac{1}{2}\left(1-c_\alpha \right) 
		     \left( 1\pm \cos\theta \right)	\nonumber \\
	\Cov_\pm(E) &=& \frac{1}{2}\left(1-c_\alpha \right) 
		\left( 1 \mp \frac{T-2\nu}{l} \right).	\nonumber
\eea
We refer the reader to ref.~\cite{BKS-MP} for radiative corrections
(RC) to unpolarized neutrino--electron scattering. RC are not included
in the results of the present paper.

If we sum over the helicities of the final-state electron,
\eq{fullpolarNu} immediately leads us to (the superscript $s$ indicates
the sum over the helicities)
\bea
        \left[\frac{d\sigma}{dE}\right]_{\nu}^{\hat{n},s}
	 &=& \frac{2mG_{\mu}^2}{\pi}
        \left[ \left( 1+ c_\alpha \right) 
                \left(\,\gl^2 -\gl \gr \frac{mz}{\nu}\right)
                        \right. \nonumber  \\
        & & \left. +\;\;\gr^2 \left(1-z\right)^2\left(
              1-c_\alpha + \frac{m z c_\alpha}{\nu-T}
				\right) \right],
\label{eq:polarNu}
\eea 
in agreement with ref.~\cite{RS}. As we can see from
\eq{fullpolarANu}, the corresponding formula for
antineutrino--electron scattering is simply obtained by interchanging
$\gl$ and $\gr$ and inverting the sign of $c_\alpha$ in
\eq{polarNu}. Furthermore, if the initial-state electrons are not
polarized, then we must also average over their polarizations, thus
obtaining the well-known SM result for the unpolarized $\nu_l-e^-$
elastic scattering computed long ago by 't~Hooft \cite{tH} (the
superscript $a$ indicates the average over the initial polarizations
of the electron)
\be
   \left[\frac{d\sigma}{dE}\right]_{\nu}^{a,s} = \;
	\frac{2mG_{\mu}^2}{\pi} \left[\, \gl^2 +\gr^2 
   	\left(1-z\right)^2 -\gl \gr 
        \frac{m z}{\nu}\right].
\label{eq:unpolar}
\ee
Once again, for $\nuov_l-e^-$ scattering we should just interchange
$\gl$ and $\gr$.

\begin{figure}[tbp]
\vspace{-6cm}\hspace{-4cm}\includegraphics[width=21cm]{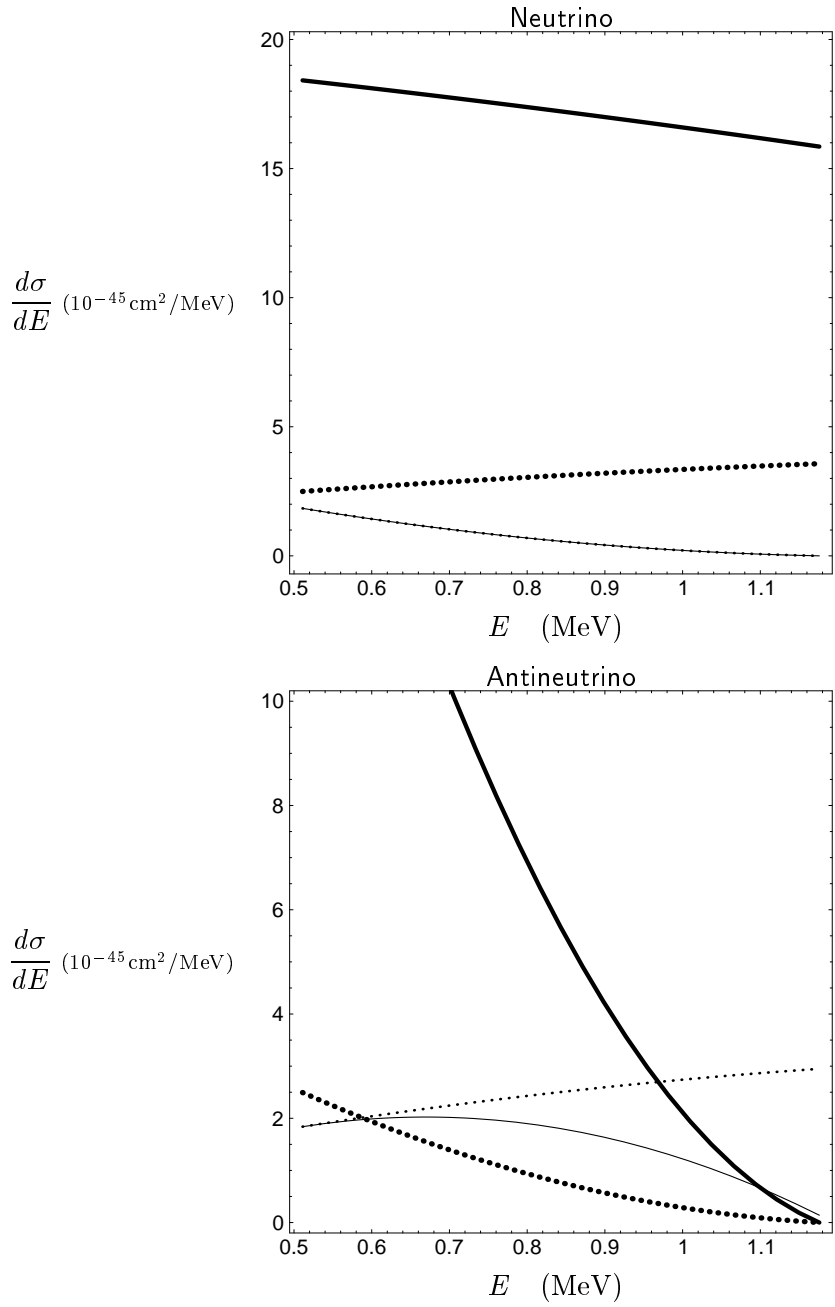}
\vspace{-7cm}
\caption{\sf Polarized differential cross sections for $\nu_l-e$
(upper panel) and $\nuov_l-e$ (lower panel) elastic
scatterings. $\nu=0.862\mev$. Solid (dotted) lines indicate $\nu_e$
($\nu_{\mu,\tau}$). Thick and thin lines represent, respectively,
$c_{\alpha}=+1$ and $c_{\alpha}=-1$.}
\label{figure:diffsigma}
\end{figure}

Irrespective of the final helicity of the electron, we note that when
the initial-state electron is polarized towards the incoming neutrino
(i.e.~when $c_{\alpha}=-1$), the differential cross section
\eq{fullpolarNu} does not depend on $\gl$ and is therefore independent
of the neutrino flavor. If, on the contrary, we reverse the
polarization of the initial-state electron (i.e.~when
$c_{\alpha}=+1$), \eq{fullpolarNu} depends both on $\gl$ and on $\gr$,
and is different for $\nu_e$ and $\nu_{\mu,\tau}$. This can be seen in
the upper panel of \fig{diffsigma}, where we have plotted the
differential cross sections for polarized $\nu_l-e^-$ elastic
scattering for an incident neutrino energy $\nu=0.862 \mev$, summing
over the helicities of the final-state electron. (This value of $\nu$
was chosen for its relevance in the study of solar neutrinos:
$\nu=0.862\mev$ is the energy of the monochromatic neutrinos produced
by electron capture on \BeSeven in the solar interior.) Solid (dotted)
lines indicate $\nu_e$ ($\nu_{\mu,\tau}$) and their thickness
represents the two different initial-state electron polarizations
$c_{\alpha}=+1$ and $c_{\alpha}=-1$ (thick and thin,
respectively). The thin solid and thin dotted curves overlap, while
the thick ones differ significantly. The lower panel of
\fig{diffsigma} shows the corresponding differential cross sections
for polarized $\nuov_l-e^-$ elastic scattering, with the same value
$\nu=0.862\mev$ for reasons of comparison.

Lead by these simple considerations, we introduce
$P_{f}^{h}(\nu,c_\alpha)$, the polarization asymmetry
defined as
\be
	P_{f}^{h}(\nu,c_\alpha) \; \; = \; \; \frac{
{\displaystyle 	\left[\sigma\right]^{\hat{n},h}_f -
		\left[\sigma\right]^{-\hat{n},h}_f}}{
{\displaystyle	\left[\sigma\right]^{\hat{n},h}_f +
		\left[\sigma\right]^{-\hat{n},h}_f}} \; , 
\label{eq:P}
\ee
where the superscript $h$ can be either $\pm$ or $s$ (we remind the
reader that the superscript $\pm$ indicates positive or negative
helicity of the final-state electron, while $s$ denotes their sum),
$f=\nu_l$ or $\nuov_l$ ($l=e$, $\mu$ or $\tau$), and
$\left[\sigma\right]^{\hat{n},h}_f$ is the total cross section
\be
	\left[\sigma\right]^{\hat{n},h}_f  \; = \;
\int_m^{E_\smallmax} 
\left[\frac{d\sigma}{dE}\right]^{\hat{n},h}_f dE
\label{eq:Sigma}
\ee
($[\sigma]^{-\hat{n},h}_f$ is the corresponding total cross section
when the initial electron polarization vector lies in the direction
{\boldmath{$-\hat{n}$}}).  Similarly, employing differential cross
sections rather than total ones, one can also define the polarization
asymmetry parameter $p_{f}^{h}(\nu,E,c_\alpha)$.

\begin{figure}[tbp]
\vspace{-6cm}\hspace{-4cm}\includegraphics[width=21cm]{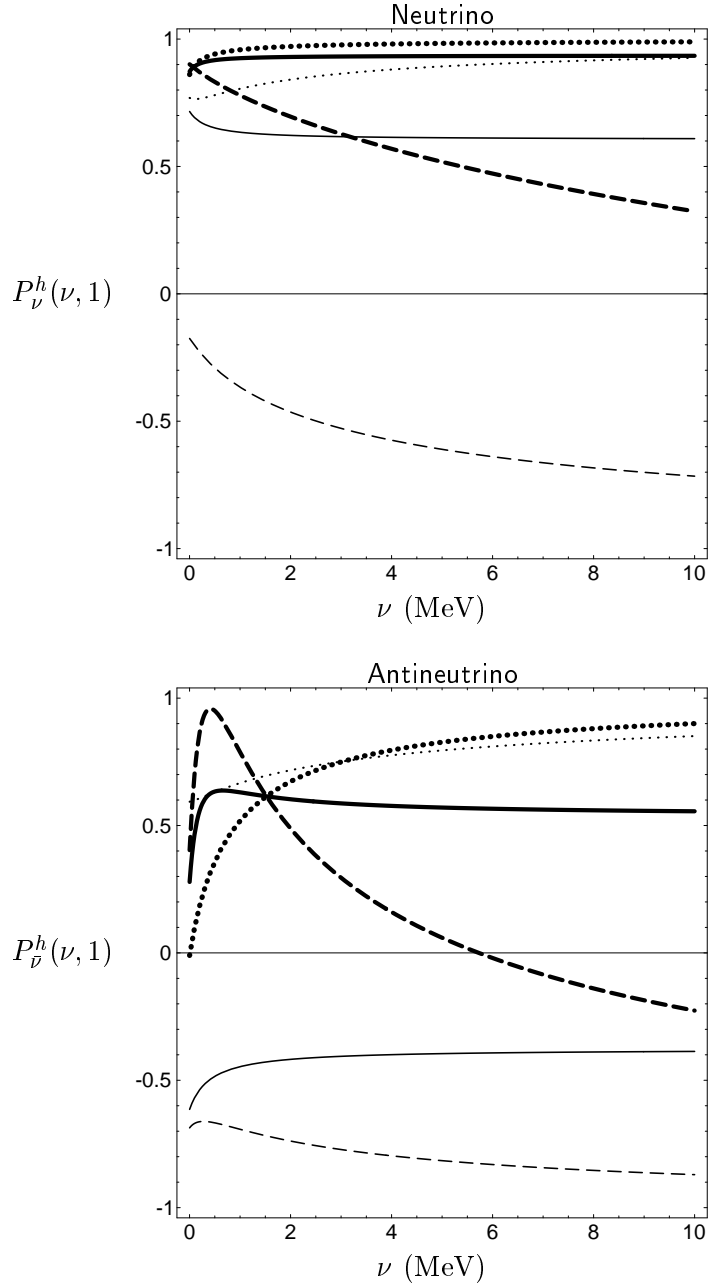}
\vspace{-7.5cm}
\caption{\sf $P_{f}^{h}(\nu,c_\alpha\!\!=\!\!1)$ for incident
neutrinos (upper panel) and antineutrinos (lower panel) of energy $\nu
\in [1\kev, 10\mev]$. Thick (thin) lines correspond to $\nu_e$ and
$\nuov_e$ ($\nu_{\mu,\tau}$ and $\nuov_{\mu,\tau}$).  Dashed (dotted)
lines represent polarization asymmetries for positive (negative)
helicity of the final-state electron, while solid curves indicate 
helicity sums.}
\label{figure:PNA}
\end{figure}

The polarization asymmetry provides a sensitive tool to investigate
the flavor of the incoming (anti)neutrinos.  In the upper panel of
\fig{PNA} we plotted $P_{f}^{h}(\nu,c_\alpha)$ for incident neutrinos
of energy $\nu \in [1\kev, 10\mev]$ and $c_\alpha=1$; the optimal
value $c_\alpha=1$ was chosen to maximize the flavor sensitivity of
the asymmetry parameter. In this panel, the three thick (thin) lines
indicate $\nu_e$ ($\nu_{\mu,\tau}$); plots for positive and negative
helicity of the recoil electron have been drawn dashed and dotted,
respectively, while solid lines have been employed for final helicity
sums. If we focus on the solid lines, we notice their dependence on
the incident neutrino flavor. We also note that, for neutrino
scattering, the dependence on the flavor would be increased if the
final helicity could be measured, and attention limited to events with
positive helicity (upper panel, dashed lines). The lower panel of
fig.~2 shows the corresponding plots of $P_{f}^{h}(\nu,1)$ for
antineutrinos.

The flavor composition of a $\nu_l$ (or $\nuov_l$) beam can be
investigated in a flux-independent way by measuring the number of
scattering events off electrons polarized along and opposite to the
direction of the incoming neutrino (or antineutrino) momentum.  The
comparison of this experimental measurement with the SM prediction
[obtained via eqs.~(\ref{eq:P}) and (\ref{eq:Sigma})] provides us with
the flavor content of the incident $\nu_l$ (or $\nuov_l$) beam,
i.e.~the fraction of $\nu_e$ ($\nuov_e$) vs.~$\nu_{\mu,\tau}$
($\nuov_{\mu,\tau}$) in the incident flux. Conversely, if the incoming
(anti)neutrino flavor is known, this comparison provides the fraction
of neutrinos vs.~antineutrinos.

The use of the polarization asymmetry parameter for
(anti)neutrino--electron scattering was advocated long ago to test the
predictions of different models for a unified theory of
electromagnetic and weak interactions \cite{tg}. More recently, the
scattering of $\nu_e$ and $\nuov_e$ on a polarized electron target has
been suggested as a test of the neutrino magnetic moment
\cite{RS}. The analyses of this note provide new and explicit
motivation for the idea of scattering (anti)neutrinos off polarized
electrons.

%%%%%%%%%%%%%%%%%%%%%%%%%%%%%%% ACKNOWLEDG %%%%%%%%%%%%%%%%%%%%%%%%%%%%
%\bigskip
It is a pleasure to thank
C.~Broggini, R.~S.~Raghavan, A.~Rossi and A.~Sirlin for
very useful comments and discussions. 

%%%%%%%%%%%%%%%%%%%%%%%%%%%%%%% REFERENCES %%%%%%%%%%%%%%%%%%%%%%%%%%%%

%%%%%%%%%%%%%%%%%%%%%%%%%%%%%%% FIGURES %%%%%%%%%%%%%%%%%%%%%%%%%%%%%%%

\end{document}